\documentclass[traditabstract]{aa} % for the abstract without structuration

\usepackage{graphicx}
\usepackage{natbib}
\usepackage{float}
\usepackage{placeins}
\usepackage{caption}
\usepackage[absolute, showboxes]{textpos} %\usepackage[absolute, showboxes, overlay]{textpos}
\usepackage{txfonts}
\usepackage{color}
\newcommand{\ecs}{erg cm$^{-3}$ s$^{-1}$~}
\newcommand{\inten}{erg cm$^{-2}$ s$^{-1}$~}

\newcommand{\cmt}{cm$^{-3}$~}

\newcommand{\gm}{$\Gamma_{\rm mech}$~}

\newcommand{\go}{$G_0$~}

\newcommand{\hcop}{${\rm HCO}^+$~}
\newcommand{\thco}{$^{13}$CO~}

\newcommand{\Msun}{${\rm M}_{\odot}$}

\begin{document}
\title{Constraining cloud parameters using high density gas tracers in galaxies}

   \subtitle{}

   \author{M. V. Kazandjian\inst{1}, 
     I. Pelupessy\inst{1,2}, 
     R. Meijerink\inst{1}, 
     F. P. Israel\inst{1},
     C. M. Coppola\inst{3},
     M. J. F. Rosenberg\inst{1}, 
     M. Spaans\inst{4}}

   \institute{
   Leiden Observatory, Leiden University, P.O. Box 9513, 2300 RA Leiden,
   The Netherlands\\
   \email{mher@strw.leidenuniv.nl}
   \and
   Institute for Marine and Atmospheric research Utrecht, Utrecht University,
   Princetonplein 5, 3584 CC Utrecht, The Netherlands
   \and
   Dipartimento di Chimica, Universit\'{a} degli Studi di Bari, Via Orabona 4,
   I-70126 Bari, Italy
   \and
   Kapteyn Astronomical Institute, PO Box 800, 9700 AV Groningen, The
   Netherlands}

   \date{Received July 7, 2014; accepted May 24, 2016}

   \abstract {
Far-infrared molecular emission is an important tool used to understand the
excitation mechanisms of the gas in the inter-stellar medium of star-forming
galaxies.  In the present work, we model the emission from rotational
transitions with critical
densities $n \gtrsim 10^4$~\cmt. We include $ 4-3 < J \le 15-14$ transitions of
CO and \thco, in addition to $J \le 7-6$ transitions of HCN, HNC, and \hcop on
galactic scales. We do this by re-sampling high density gas in a hydrodynamic
model of a gas-rich disk galaxy, assuming that the density field of the
interstellar medium of the model galaxy follows the probability density function
(PDF) inferred from the resolved low density scales.
We find that in a narrow gas density PDF, with a mean density of $\sim 10$~\cmt~
and a dispersion $\sigma = 2.1$ in the log of the density, most of the emission
of molecular lines, even of gas with critical densities $> 10^4$~\cmt, emanates
from the 10-1000~\cmt~part of the PDF. We construct synthetic emission maps for
the central 2 kpc of the galaxy and fit the line ratios of CO and \thco~ up to
$J = 15-14$, as well as HCN, HNC, and \hcop up to $J = 7-6$, using one
photo-dissociation region (PDR) model.  We attribute the goodness of the one
component fits for our model galaxy to the fact that the distribution of the
luminosity, as a function of density, is peaked at gas densities between 10 and
1000~\cmt, with negligible contribution from denser gas.  Specifically, the Mach
number, $\mathcal{M}$, of the model galaxy is $\sim 10$.

We explore the impact of different log-normal density PDFs on the distribution
of the line-luminosity as a function of density, and we show that it is necessary to
have a broad dispersion, corresponding to Mach numbers $\gtrsim 30$ in order to
obtain significant ($ > 10$\%) emission from $n > 10^4$~\cmt~ gas. Such Mach
numbers are expected in star-forming galaxies, luminous infrared galaxies
(LIRGS), and ultra-luminous infrared galaxies (ULIRGS).
This method provides a way to constrain the global PDF of the ISM of galaxies
from observations of molecular line emission. As an example, by fitting line
ratios of HCN(1-0), HNC(1-0), and \hcop(1-0) for a sample of LIRGS and ULIRGS
using mechanically heated PDRs, we constrain the Mach number of these galaxies
to $29 <\mathcal{M} < 77$.}


\authorrunning{Kazandjian {et al.}}
\titlerunning{Constraining cloud parameters in galaxy centers}
\maketitle

\section{Introduction} \label{sec:paper4_intro}

The study of the distribution of molecular gas in star-forming galaxies provides
us with an understanding of star formation processes and their relation to galactic
evolution.
In these studies carbon monoxide (CO) is used as a tracer of star-forming
regions and dust, since in these cold regions ($ T < 100$~K) H$_2$ is virtually
invisible.
 The various rotational transitions of CO emit in the far-infrared (FIR)
spectrum, and are able to penetrate deep into clouds with high column densities,
which are otherwise opaque to visible light.
CO lines are usually optically thick and their emission emanates from the
C$^+$/C/CO transition zone \citep{Wolfire89-1}, with a small contribution to the
intensity from the deeper part of the cloud \citep{meijerink2007-1}. On the
other hand, other molecules, whose emission lines are optically thin beacuse of their
lower column densities, probe greater depths of the cloud compared to CO.  These
less abundant molecules (e.g., \thco) have a weaker signal than CO, and a longer
integration time is required in-order to observe them.  Since ALMA became
available, it has become possible to obtain well-resolved molecular emission
maps of star-forming galaxies in the local universe, due to its high
sensitivity, spatial and spectral resolution.  In particular, many species have
been observed with ALMA, including the ones we consider in this paper, namely
CO, \thco, HCN, HNC, and \hcop~ \citep[e.g.,][]{Imanishi13-1, Saito13-1,
Combes13-1, Scoville13-1, Combes14-1}.

Massive stars play an important role in the dynamics of the gas around the
region in which they form.  Although the number of massive stars ($ M >
10$~\Msun) is about 0.1\% of the total stellar population, they emit more than
99\% of the total ultraviolet (FUV) radiation.
This FUV radiation is one of the main heating mechanisms in the ISM of star-forming
regions. Such regions are referred to as photon-dominated regions (PDRs)
and they have been studied since the 1980s \citep{tielenshb1985,
Hollenbach1999}.
Since then, our knowledge of the chemical and thermal properties of these
regions has been improving.
Since molecular clouds are almost invariably accompanied by young luminous
stars, most of the molecular ISM forms in the FUV shielded region of a PDR, and
thus this is the environment where the formation of CO and other molecular 
species can be studied.
In addition, the life span of massive stars is short, on the order of 10 Myrs,
thus they are the only ones that detonate as supernovae, liberating a
significant amount of energy into their surroundings and perturbing it.  A small
fraction of the energy is re-absorbed into the ISM, which heats up the gas
\citep{Usero07-1, falgaron2007p, loenen2008}.  In addition, starbursts (SB)
occur in centers of galaxies, where the molecular ISM can be affected by X-rays
of an accreting black hole (AGN) and enhanced cosmic ray rates or shocks
\citep[][among many others]{maloney96, komossa03, martin06-1, oka2005ApJ_1,
vandertak2006_1, pan2009-1, papadopoulos10, meijerink11, meijerink13-1,
Rosenberg14} that ionize and heat the gas.

By constructing numerical models of such regions, the various heating mechanisms
can be identified.  However, there is no consensus about which combination of
lines define a strong diagnostic of the different heating mechanisms.  This is
mainly due to the lack of extensive data which would probe the various
components of star-forming regions in extra-galactic sources.  Direct and
self-consistent modeling of the hydrodynamics, radiative transfer and chemistry
at the galaxy scale is computationally challenging, thus some simplifying
approximations are usually employed. In the simplest case it is commonly assumed
that the gas has uniform properties, or is composed of a small number of uniform
components.
In reality, on the scale of a galaxy or on the kpc scale of starbursting
regions, the gas density follows a continuous distribution.  Although the exact
functional form of this distribution is currently under debate
\citep[e.g.][]{Nordlund99-1}, it is believed that in SB regions, where the gas
is thought to be supersonically turbulent \citep{norman96-1, Goldman12-1} the
density distribution of the gas is a log-normal function \citep{Vazquez94-1,
Nordlund99-1, wada01-a, wada07-1, kritsuk11-1, ballesterosParedes11-1,
Burkhart13-1, hopkins13-1}.
This is a universal result, independent of scale and spatial location, although
the mean and the dispersion can vary spatially.
Self-gravitating clouds can add a power-law tail to the density PDF
\citep{kainulainen09-1,froebrich10-1,russeil13-1,AlvesdeOliveira14-1,Schneider14-1}.
However, \cite{kainulainen2013-1} claim that such gravitational effects are
negligible on the scale of giant molecular clouds, where the molecular emission
we are interested in emanates.

In \cite{mvk15-a}, we studied the effect of mechanical heating (\gm~hereafter)
on molecular emission grids and identified some diagnostic line ratios to
constrain cloud parameters including mechanical heating.  For example, we showed
that low-$J$ to high-$J$ intensity ratios of high density tracers will yield a
good estimate of the mechanical heating rate.  In \cite{mvk15-b}, KP15b
hereafter, we applied the models by \cite{mvk15-b} to realistic models of the
ISM taken from simulations of quiescent dwarf and disk galaxies by
\cite{inti2009-1}.  We showed that it is possible to constrain mechanical
heating just using $J < 4-3$ CO and \thco~ line intensity ratio from ground
based observations.
This is consistent with the suggestion by \cite{israel03} and
\cite{israel2009-1} that shock heating is necessary to interpret the high
excitation of CO and \thco~in star-forming galaxies. This was later verified by
\cite{loenen2008}, where it was shown that mechanically heated PDR models are
necessary to fit the line ratios of molecular emission of high density tracers
in such systems.

Following up on the work done by KP15b, we include high density gas ($n >
10^4$~\cmt) to produce more realistic synthetic emission maps of a simulated
disk-like galaxy, thus accounting for the contribution of this dense gas to the
molecular line emission. This is not trivial as global, galaxy wide models of
the star-forming ISM are constrained by the finite resolution of the simulations
in the density they can probe.
  This paper is divided into two main parts.  In the first part,
we present a new method to incorporate high density gas to account for its
contribution to the emission of the high density tracers, employing the
plausible assumption, on theoretical grounds \citep[e.g.][]{Nordlund99-1} that
the density field follows a log-normal distribution.
In the methods section, we describe the procedure with which the sampling of the
high density gas is accomplished. Once we have derived a re-sampled density
field we can employ the same procedure as in KP15b to model the  line emission
of molecular species.  In Section-\ref{subsec:paper4_emissionmaps}, we highlight
the main trends in the emission of the $J = 5-4$ to $J = 15-14$ transitions of
CO and \thco~ tracing the densest gas, along with the line emission of high
density tracers HCN, HNC and \hcop for transitions up to $J = 7-6$.  In
Section-\ref{subsec:paper4_constrainig}, we fit emission line ratios using a
mechanically heated PDR (mPDR hereafter) and constrain the gas parameters of the
model disk galaxy.  In the second part of the paper,
Section-\ref{section:constrainingpdf}, we will follow the reverse path and
examine what constraints can be placed on the PDF from molecular line emissions,
following the same modeling  approach as in the first part. We discuss the
effect of the shape and width of the different density profiles on the emission
of high density tracers. In particular, we discuss the possibility of
constraining the dispersion and the mean of an assumed log-normal density
distribution using line ratios of high density tracers.  We finalize with a
summary and general remarks.

\section{Methods \label{sec:paper4_methods}}

The numerical methods we implement in this paper are similar to those in KP15b.
We will focus exclusively on a single model disk galaxy, but the methods
developed here could be applied to other models. We implement a recipe for the
introduction of high density gas $n > 10^4$~\cmt, which is necessary to model
the emission of molecular lines with critical densities\footnote{We use the
following definition of the critical density, $n_{crit} \equiv k_{ij} / A_{ji}$,
where $k_{ij}$ is the collision coefficient from the level $i$ to the level $j$
and $A_{ji}$ is the Einstein coefficient (of spontaneous decay from level $j$ to
level $i$.} ($n_{\rm crit}$) in the range of $10^4 - 10^8$~\cmt .  In the
following, we summarize briefly the methods used in KP15b.

\subsection{Galaxy model, radiative transfer and subgrid modelling}

The model galaxy we use in this paper is the disk galaxy simulated by
\cite{inti2009-1}, with a total mass of $10^{10}$~\Msun~and a gas content
representing 10\% of the total mass. This represents a typical example of a
quiescent star-forming galaxy. The code {\sc Fi} \citep{inti2009-1} was used to
evolve the galaxy to dynamical equilibrium (in total for $\sim 1$ Gyr).
The simulation code implements an Oct-tree algorithm which is used to compute
the self-gravity \citep{barnes86} and a TreeSPH method for evolving the
hydrodynamics \citep{monaghan92, springelHernquist02}, with a recipe for star
formation based on the Jeans mass criterion \citep{intiPhdT}.  The adopted ISM
model is based on the simplified model by \cite{wolfire03}, where the local FUV
field in the neighborhood of the SPH particles is calculated using the
distribution of stellar sources and population synthesis models by
\cite{bruzualCharlot93, bruzualCharlot03} and \cite{parravano2003}.  The local
average mechanical heating rate due to the self-interacting SPH particles is
derived from the prescription by \cite{springel05-1}: the local mechanical
heating rate is estimated using the local dissipation by the artificial
viscosity terms, which in this model ultimately derive mainly from the localized
supernova heating.  For the work presented here the most important feature of
these simulations is that they provide us with the information necessary for
further subgrid modeling in post-processing mode. Details about this procedure
can be found in KP15b.  In particular, the gas density $n$, the mean local
mechanical heating rate \gm, the local FUV flux $G$ (measured in units of \go $= 1.6\times 10^{-3}$\inten) and the mean $A_V$ of the
SPH particles are used. We refer to these as the main physical parameters of the
gas, which are essential to parametrize the state of PDRs that are used, to
obtain the emission maps.  This method of computing the emission can be
generally applied to other simulations, as long as they provide these physical
parameters.

The PDR modeling \citep{meijerink2005-1} consists of a comprehensive set of
chemical reactions between the species of the chemical network by
\cite{umist1999}.  The main assumption in the extended subgrid modeling, is
that these PDRs are in thermal and chemical equilibrium.  We post-process an
equilibrium snapshot of the SPH simulation by applying these PDR models to the
local conditions sampled by each particle, and estimate the column densities of
the molecules, abundances of the colliding species, and the mean gas temperature
of the molecular clouds.  These are the main ingredients necessary to compute
the emission emanating from an SPH particle.  PDR models are non-homogeneous by
definition, as there are steep gradients in the kinetic temperature and the
abundances of chemical species.  By assuming the large velocity gradient (LVG)
approximation \citep{sobolev1960} using {\sc Radex} \citep{radex2}, weighted
quantities from the PDR models were used as input to Radex to compute the
emission of the molecular species studied in this paper \citep{mvk15-a}.  For
more details on computing the emission and constructing the emission maps we
refer to Section-2.3 of KP15b, where this method can be applied to any SPH
simulation of a star-forming galaxy that provides the ingredients mentioned
above.  For the work in this paper, we do not include any AGN or enhanced cosmic
ray physics, since these are not relevant for the simulation we have used.  XDR
and or enhanced cosmic ray models are necessary in modeling the ISM of ULIRGS,
as was discussed by, e.g., \cite{papadopoulos10} and \cite{meijerink11} and
shown in the application of the models to Arp 299 by \cite{2014A&A...568A..90R}.

\subsection{Sampling the high-density gas}

One of the main purposes of this paper is to constrain cloud parameters using
the molecular transitions mentioned earlier.  The typical critical densities of
these transitions are between $\sim 10^5$ and $\sim 10^8$~\cmt.  The highest
density reached in the SPH simulation is $~10^4$~\cmt, thus the rotational
levels we are interested in are subthermally populated in the non-LTE (local
thermal equilibrium) regime. Therefore the intensity of the emission associated
with these transitions is weak.  In-order to make a more realistic
representation of the molecular line emission of the ISM of the simulated
galaxy, we resort to a recipe to sample particles up to $10^6$~\cmt.  In what
follows, we describe the prescription we have used to sample particles to such
densities.

The sampling scheme we adopt is based on the assumption that the gas density PDF
of the cold neutral medium (CNM) and the molecular gas is a log-normal function,
given by Eq-\ref{eq:paper4_ln-func} \citep{nrcpp08}:

%---------------
\begin{equation} \label{eq:paper4_ln-func}
  \frac{dp}{d \ln{n}} = \frac{1}{\sqrt{2 \pi \sigma^2}} \exp{ \left( - \frac{1}{2} \left[ \frac{\ln{n} - \mu }{\sigma} \right]^2 \right) }
\end{equation}
%---------------

\noindent where $\mu$ is the natural logarithm ($\ln$) of the median density
($n_{\rm med}$), and $\sigma$ is the width of the log-normal distribution. Such
PDFs are expected in the inner ($\sim$~kpc) ISM of galaxies.  For example,
simulations by \cite{wada01-a} and \cite{wada01-b} reveal that the PDF of the
gas density is log-normally distributed over seven orders of magnitude for
densities ranging from $\sim 4$~\cmt~to $\sim 4 \times 10^7$~\cmt.  Log-normal
density distributions have been discussed by other groups as well, and we refer
the reader to Section-\ref{sec:paper4_intro} for more references.  The
simulations by Wada are evolved using a grid based code with a box size $\sim
1$~kpc and include AGN feedback, whereas our simulation was evolved with an  SPH
code and without AGN feedback with a scale length of $\sim 10$~kpc.  The major
assumption in our sampling is that the gas of our model disk galaxy follows a
log-normal distribution for $n > 10^{-2}$~\cmt.  Using this assumption we can
sample and add high density gas beyond the maximum of $10^4$~\cmt~of the
simulated galaxy.
Of course such re-sampling does not provide a realistic spatial distribution of
high density gas, as the re-sampled particles are assumed to be placed randomly
within the smoothing length of SPH particles that are dense enough to support
star formation. We can use this simplification as long as we do not try to
construct maps resolving scales smaller than the original spatial resolution of
$\sim$ 0.05 kpc.

\begin{figure}[!th]
\begin{minipage}[b]{1.0\linewidth} 
\centering
\includegraphics[scale=1.0]{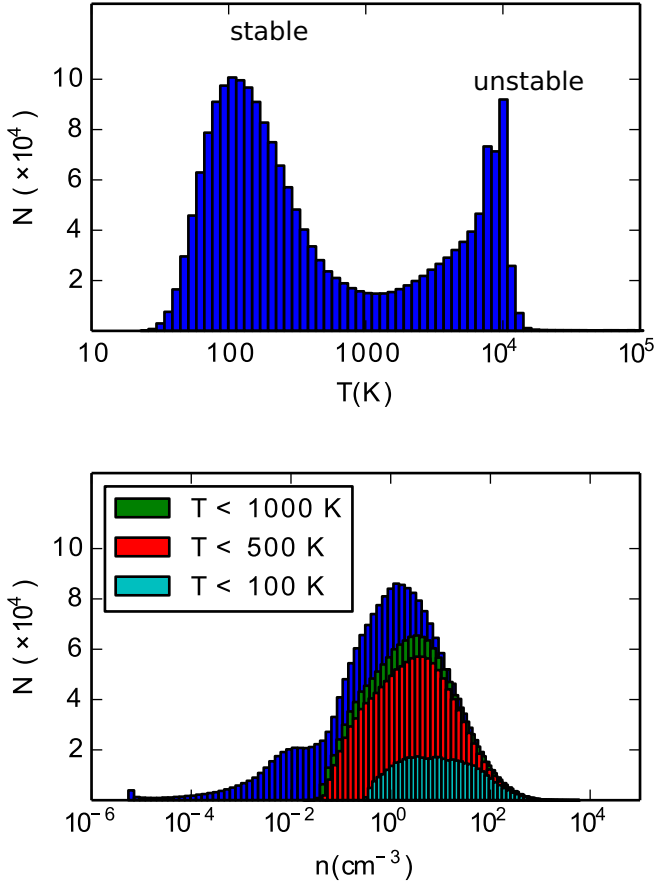}
\end{minipage}
\caption{ (Top) Histogram of the kinetic temperature of the SPH particles. 
(Bottom) Histogram of the gas density of the SPH particles.  In this same panel,
the histograms for the gas densities of sub-populations of the gas are also
shown.  The red, green and cyan histograms correspond to gas particles with
temperature below $T = 1000, 500, 100$~K respectively.  These sub-populations of
the gas particles are thermally stable corresponding  to the peak around $T =
300$~K in the top panel.  The vertical axis refers to the number of SPH
particles within each bin.
\label{fig:paper4_T-n-hist}} 
\end{figure}

In Figure \ref{fig:paper4_T-n-hist}, we show the distributions of the gas
density of our simulation along with that of the temperature.  The SPH
simulation consists of $N = 2 \times 10^6$ particles.  The distribution of the
temperature has a peak at low temperature and a peak at high temperature,
similar to the PDF in Figure 4 in \cite{wada01-b}.  The low temperature peak
around $T = 300$ K corresponds to gas that is thermally stable, whereas the
higher temperature peak around $T = 10000$ K corresponds to gas that is
thermally unstable \citep{Schwarz1972-1}.
On the other hand, the distribution of the density exhibits just one prominent
peak near $n \sim 1$~\cmt, with a small saddle-like feature at a lower density
of $\sim 10^{-2}$~\cmt~ (see the bottom panel of
Figure \ref{fig:paper4_T-n-hist}).  The histogram of the density of the stable
gas in the bottom panel, shows that the density of this gas ranges between
$10^{-2} < n < 10^4$~\cmt. The lower bound of this interval is shifted towards
higher densities as the gas populations are limited to lower temperatures,
indicating efficient cooling of the gas for high densities.  The temperature of
the SPH particles represents its peripheral temperature.  Thus, although the
temperature might be too high for the formation of molecules at the surface,
molecules could form in the PDRs present in the subgrid modeling.

\begin{figure}[!th]
\begin{minipage}[b]{1.0\linewidth} 
\centering
\includegraphics[scale=1.0]{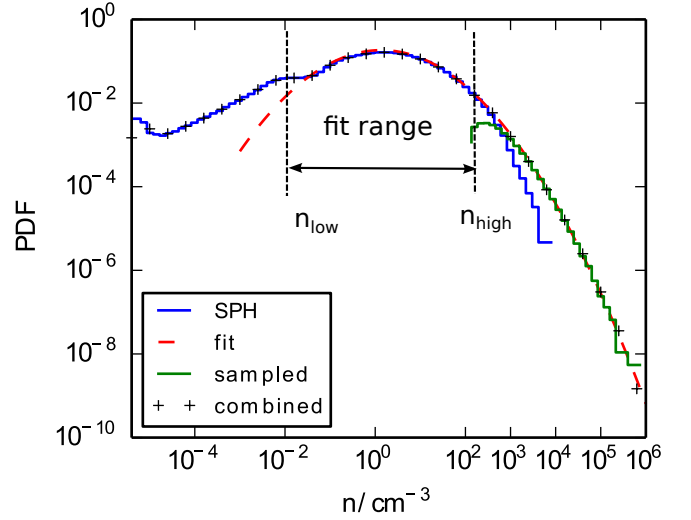}
\end{minipage}
\caption{Probability density function of the SPH particles.  The blue curve
corresponds to the PDF of gas density $n$ and it is proportional to $d N / d
\log(n)$; in other words this curve represents the probability of finding an SPH
particle within a certain interval of $\log(n)$.  The dashed-red curve is the
log-normal fit of the PDF of the blue curve in the range $10^{-2}$ to $10^2$. 
The green curve is the PDF of the sampled population from the original SPH
particles while keeping the samples with $n > 10^2$~\cmt.
The black crosses trace the combined PDF of the sampled and the original set of
the SPH particles. The median density and the dispersion are $n_{\rm med} =
1.3$~\cmt~ and $\sigma = 2.1$ respectively.
\label{fig:paper4_sampling-fit}} 
\end{figure}

The 1000 K mark seems to be a natural boundary between the thermally stable and
the unstable population (see top panel of Figure \ref{fig:paper4_T-n-hist}).
Hence we use the minimum density range for this population, $n \sim
10^{-2}$~\cmt, as the lower limit for the fit of the PDF that we apply.  This is
depicted in Figure \ref{fig:paper4_sampling-fit}, where we fit a log-normal
function to the gas density around the peak at $n = 1$~\cmt.
The density range of the fit is $n_{\rm low} = 10^{-2} < n < n_{high} =
10^2$~\cmt.
To find the best fit of the actual density PDF, we examined other values for
$n_{\rm low}$ and $n_{high}$.  For example, choosing $n_{\rm low} =
10^{-1}$~\cmt~ causes the fit PDF to drop faster than the original distribution
for $n > 10^2$~\cmt, which reduces the probability of sampling particles with $n
> 10^5$~\cmt~ to less than one particle in a sample of one billion.  On the
other hand, setting $n_{high} = 10^3$~\cmt~ does not affect the outcome of the
fit.  Using the best fit density PDF, we notice that only the gas within the
$10^{-2} < n < 10^2$~\cmt~ range is log-normally distributed, where the original
distribution starts deviating from the log-normal fit for $n > 10^2$~\cmt~
(compare blue and red curves in Figure \ref{fig:paper4_sampling-fit}).  The
deviation from a log-normal distribution is likely due to the resolution limit
of the SPH simulation.
For example, the PDF of a simulation with $4 \times 10^5$ SPH particles start to
deviate from log-normality for $n > 10^{1.5}$~\cmt.
A similar deviation in the density distribution fit by \cite{wada01-b} is
observed in the $n \gtrsim 10^6$~\cmt~ range, most likely due to similar
numerical
resolution constraints.  Here we will assume that the gas density will be
distributed according to a log-normal function in the infinite $N$-limit.

For each SPH particle $i$, parent particle hereafter, we sample from the fit PDF
an ensemble $\{n_s\}$ of 100 sub-particles within the smoothing length of the
parent, where $n_s$ and $n_i$ are the gas number densities of the sampled and
parent particles respectively.  The ensemble $\{n_s\}$ is chosen such that $n_s
> n_i$, which enforces having sampled particles that are denser than the parent.
The particles with $n > 10^2$~\cmt~ constitute 2\% of the gas mass, as their
total number is $\sim 40,000$.  The PDF of  the sampled particles is shown in
green in Figure \ref{fig:paper4_sampling-fit}, where the sampled PDF follows the
fit log-normal distribution very accurately for densities $n > 10^3$~\cmt.  The
bias of sampling particles with $n_s > n_i$ affects the shape of the
distribution between $ n = 10^2$~\cmt~ to $10^3$~\cmt.  This discrepancy is
resolved by adjusting the weights of the sampled and parent particles along bins
in density range such that the new weights match the fit PDF.
These weights are used to adjust the masses of the parent and sampled particles,
which ensures that the total mass of the system is conserved.  The combined PDF
is overlaid as black crosses in Figure \ref{fig:paper4_sampling-fit}.

The re-sampling is restricted to particles with $n_i > 10^2$~\cmt~to ensure that
the sub-particles are sampled in regions that are likely to support
star formation.  Thus, most of the far-IR molecular emission results from this
density range.  In our sampling, we ignore any spatial dependence on the gas
density distribution, where the PDFs correspond to that of all the particles in
the simulation.  Sampling new gas particles with $n > 10^2$~\cmt~ ensures that
these ensembles will lie in regions tracing CO and consequently H$_2$.  At the
outskirts of the galaxy, the H$_2$ column density is at least 100 times lower
compared to the center of the galaxy.  Moreover, since the density PDF is
log-normal, the number of SPH particles with $n > 10^4$~\cmt~in the outskirts
are so small that no emission from high density tracers is expected.
Thus, we would not have enhanced molecular emission due to the sampling, e.g.
for the high density tracers, at the edge of the galaxy, where it is not
expected to be observed.  To address this point, we examine the shape of the gas
density PDF for increasing galactocentric distances. We find that the median of
the log-normal distribution is shifted to lower densities, while the dispersion
becomes narrower towards the edge of the galaxy.  Consequently, the relative
probability of finding high density gas at the outer edge of the galaxy is
reduced compared to that of the central region. In fact, the PDF in the region
within $R < 2$~kpc is closely represented by the one we have shown in
Figure \ref{fig:paper4_sampling-fit}.  The dispersion of the density PDF for
distances $R < 2$~kpc from the center is $\sigma = 2.05$ is very close to
$\sigma = 2.1$ of the fit PDF in Figure \ref{fig:paper4_sampling-fit}.  On the
other hand, the dispersion is $\sigma = 1.2$ for $ 6 < R < 8$~kpc.
Thus, in this paper we focus on the central region of the galaxy, where the same
density PDF is used in sampling the gas particles within this region.
It is worth noting that after the sampling procedure, gas with $n >
10^4$~\cmt~constitutes only 0.5\% of the total gas mass. For example, this is
consistent with the estimate of the filling factor of the densest PDR component
derived from modeling the star-forming galaxy Arp 299 by
\cite{2014A&A...568A..90R}, even though our model galaxy is not necessarily
representative of Arp 299.

The log-normality feature of the density PDF reflects non-linearity in the
evolution of the gas, which we use as an argument for the cascade of turbulence
into small spatial scales where high density gas forms. Here we will take the
\gm~of the sampled particles to be the same as their parent SPH particle,
although in \cite{mvk15-a}[Figure 2] we have shown that there is some
dependence of \gm~on the gas density.  However, this dependence is not trivial
\cite[see][]{Nordlund99-1}. We note that for our models \gm~would never be the
dominant heating mechanism for high density gas ($>10^4$~\cmt~ ), while this
could change for modeling ULIRGs.
A similar assumption was made for the $A_V$ and $G$ of the sampled particles,
where the same values as that of the parent were used.  As a check we picked a
typical region in the central kpc of the galaxy and quantified the mean and the
standard deviation of \gm, $G$ and $A_V$ relative to the parent particle
compared to other gas particles within its smoothing length.  We found that the
spread in \gm~is within a factor of two, whereas this spread decreases to within
0.1 for $G$ and $A_V$.  Therefore, the assumption of using the same parameters
as the parent for the sampled ones would have a small effect on the produced
emission maps.

% -----------------------------------------------------
%-----------------------------------------------------

%-----------------------------------------------------
%-----------------------------------------------------

Various other authors have tried to overcome the resolution limit of their
simulations.  For example, the sampling procedure by \cite{Narayanan2008-1} is
motivated by observations of GMCs.  Their approach entails the sampling of GMCs
by assuming that half of the gas mass is represented by molecules. Moreover,
these GMCs are modeled as spherical and gravitationally bound with power-law
density profiles provided by \cite{rosolowsky05, blitz07} and
\cite{solomon1987-1}.
These sampling methods are similar to the work presented here, in the sense that
they make a (different) set of assumptions to probe high density gas not present
in the simulation.  Our basic assumption is a physically motivated extension of
the log-normal distribution.

\section{Modelling dense gas in galaxy disks} 

\subsection{Luminosity ladders and emission maps} \label{subsec:paper4_emissionmaps}

\begin{figure}[!th]
\begin{minipage}[b]{1.0\linewidth} 
\centering
\includegraphics[scale=1.0]{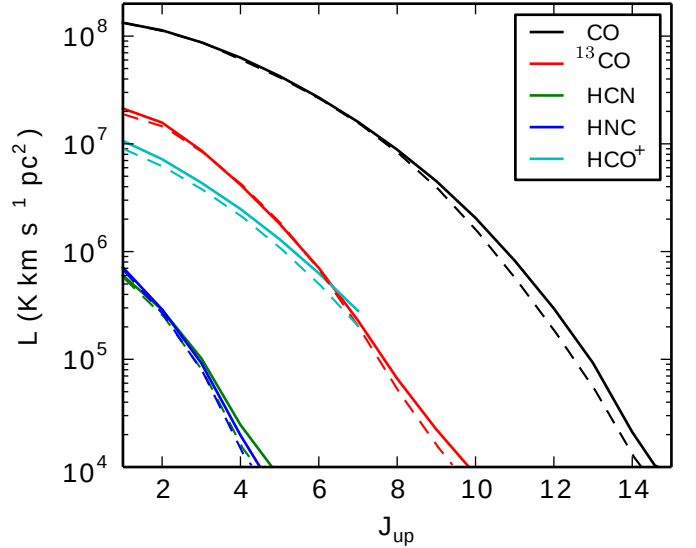}
\end{minipage}
\caption{Total luminosity of all the molecular transitions for the model galaxy
over a region of $16 \times 16$~kpc. The dashed lines correspond to the original
set of particles, whereas the solid lines represent the combined luminosity of
the sampled and the original sets.
\label{fig:paper4_total_luminosity}} 
\end{figure}

The contribution of the high density particles to the total luminosity of each
transitions is illustrated in Figure \ref{fig:paper4_total_luminosity}.  The
luminosity of a SPH particle is approximated as the product of the line flux
($F$) with the projected area $(A)$ of the SPH particle. The total luminosity of
a certain emission line of the galaxy is computed as $L = \sum_{i=1}^{i=N} F_i
A_i$, where $i$ is the index of the SPH particles and $N$ is the number of SPH
particles.  In the current implementation, the ``area'' of each sampled particle
is assumed to be $A_i / N_s$, where $N_s = 100$ is the number of the sampled
particles.   This ensures that the total area of the sampled particles adds up
to the area of the parent particle.  Such a normalization is consistent with the
SPH formalism, where all particles have the same mass.  The luminosities of the
lines due to the sampling are marginally affected, with a weak dependence on the
transition and the species.  The increase in the luminosity is due to the
dependence of the flux of the emission lines on the gas density, where in
general, the flux increases as a function of increasing density.
The dependence of the luminosity on density and its relationship to the density
PDF will be addressed in Section-\ref{section:constrainingpdf}.

\begin{figure*}[!th]
\begin{minipage}[b]{1.0\linewidth} 
\centering
\includegraphics[scale=0.68]{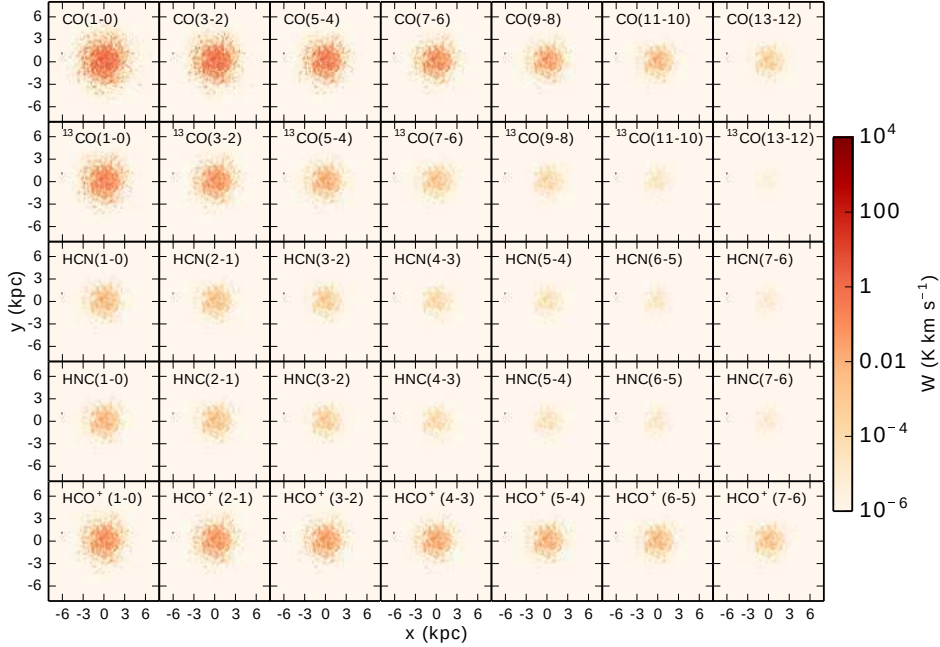}
\end{minipage}
\caption{Flux maps of the model galaxy of a selection of the CO and \thco~ lines,
and all of the transitions of HCN, HNC, \hcop.
\label{fig:paper4_flux-maps}} 
\end{figure*}

The total luminosities as a function of $J$, luminosity ``ladders'', are
computed by considering all the SPH particles within a projected box of $16
\times 16$~kpc.  In Figure \ref{fig:paper4_flux-maps} we show the flux maps of a
subset of the molecular lines.  These maps give insight on the regions of the
galaxy where most of the emission emanates.  In the next section we will use
line ratios in a similar manner to KP15b in order to constrain the gas
parameters within pixels in the central region of the galaxy.  The rotational
emission lines we have used are listed in
Table \ref{tbl:paper4_line-ratio-combinations}.

%-----------------------------------------------------
%-----------------------------------------------------

%-----------------------------------------------------
%-----------------------------------------------------

\begin{table}[h]  
\centering
\begin{tabular}{c c}
  \hline
  species & $J$-transitions\\
  \hline
  CO    & 1-0 $\rightarrow$ 15-14\\  
  \thco & 1-0 $\rightarrow$ 15-14\\  
  HCN   & 1-0 $\rightarrow$ 7-6\\  
  HNC   & 1-0 $\rightarrow$ 7-6\\  
  \hcop & 1-0 $\rightarrow$ 7-6\\  
  \hline
\end{tabular}
\caption{Rotational lines used in constructing the flux maps 
\label{tbl:paper4_line-ratio-combinations}}
\end{table}

\subsection{Constraining cloud parameters using line ratios} \label{subsec:paper4_constrainig}

With the synthetic luminosity maps at our disposal, we construct various line
ratios and use them to constrain the mechanical heating rate in addition to the
remaining gas parameters, namely $n$, $G$ and $A_V$, throughout the central  $<
2$~kpc region of the galaxy.  We follow the same approach by KP15b, where we
minimize the $\chi^2$ statistic of the line ratios of the synthetic maps against
these of a mechanically heated PDR model.

\begin{figure}[!th]
\begin{minipage}[b]{1.0\linewidth} 
\centering
\includegraphics[scale=0.6]{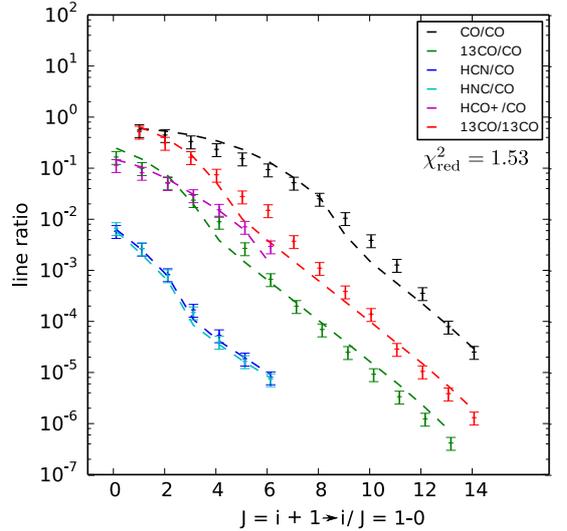}
\end{minipage}
\caption{Sample fit of the line ratios of the central pixel (0.4 x 0.4 kpc$^2$)
with a single PDR model.  The points with error bars represent the line ratios
of the specified species from the synthetic maps.  The dashed curves correspond
to the line ratios of the best fit PDR model.  All the ratios are normalized to
CO(1-0) except for the red ratios which are normalized to \thco(1-0).
\label{fig:paper4_fit1}} 
\end{figure}

We fit the cloud parameters one pixel at a time and compare them to the mean
physical parameters of the gas in that pixel.
In Figure \ref{fig:paper4_fit1}, we show a sample fit for the central pixel of
the model galaxy that has a pixel size of $0.4 \times 0.4$~kpc, which is the
same pixel size that has been assumed by KP15b.  For this fit we have considered
the luminosity ladders of CO, \thco, HCN, HNC and \hcop normalized to the
CO(1-0) transition.  In addition to these, we have included the ladder of \thco~
normalized to \thco(1-0) transitions.

%-----------------------------------------------------
%-----------------------------------------------------

The gas parameters derived from the fits for pixels of increasing distance from
the center are shown in Table \ref{tbl:paper4_fit1}.  In addition to the fit
parameters, we show the mean values of these parameters in each pixel. This
allows us to compare the fit values to the average physical conditions, which is
not possible when such fits are applied to actual observations.  We find that a
mPDR constrains the density and the mechanical heating rate to less than a
half-dex and the visual extinction to less than a factor of two.  On the other
hand, the FUV flux ($G$) is largely unconstrained.  We note that a pure PDR that
is not mechanically heated (not shown here), fails to fit most of the ratios
involving $J > 4-3$ transitions, and leads to incorrect estimates of the gas
parameters.  

\begin{table*}[!tbh]  
\centering
\begin{tabular}{c c | c c c c | c }
  \hline
$R$ &  & $\log_{10}[n]$ & $\log_{10}[G]$ & $\log_{10}[\Gamma_{\rm mech}]$ & $A_V$ &$\chi^2_{\rm red}$ \\
  \hline 
  $ < 0.4$     & mPDR   & $1.9 \pm 0.2$  & $-0.6 \pm 1.9$    & $-23.0 \pm 0.2$    & $12.2 \pm 0.4$     & $1.7 \pm 0.5$ \\
               & Actual & $1.9 \pm 0.2$  & $1.8  \pm 0.1$    & $-23.4 \pm 0.2$    & $11.2 \pm 1.3$     & --   \\
  \hline
  $ \sim 0.8$ & mPDR   & $1.9 \pm 0.3$  & $1.4 \pm 3.0$      & $-23.1 \pm 0.3$    & $12.8 \pm 2.3$     & $1.3 \pm 0.2$ \\
              & Actual & $1.8 \pm 0.1$  & $1.7 \pm 0.1$      & $-23.3 \pm 0.1$    & $9.4  \pm 1.1$     & --   \\
  \hline
  $ \sim 1.0$ & mPDR   & $1.9 \pm 0.2$  & $2.0 \pm 2.3$      & $-23.1 \pm 0.1$    & $12.0 \pm 4.4$     & $1.6 \pm 0.2$ \\
              & Actual & $1.7 \pm 0.1$  & $1.7 \pm 0.1$      & $-23.3 \pm 0.1$    & $8.5  \pm 0.6$     & --   \\
  \hline
  $ \sim 1.5$ & mPDR   & $1.7 \pm 0.4$  & $0.4 \pm 2.0$      & $-23.3 \pm 0.1$    & $8.2 \pm 1.3$      & $1.0 \pm 0.1$ \\
              & Actual & $1.5 \pm 0.1$  & $1.3 \pm 0.1$      & $-23.7 \pm 0.1$    & $5.5 \pm 0.4$      & --   \\
  \hline
\end{tabular}
\caption{Model cloud parameters fits for pixels of increasing distance $R$ (in
kpc) from the center of the galaxy.  These parameters are $n$ (in \cmt), $G$ (in
units of $G_0$), \gm~(in  \ecs) and $A_V$ (in mag).  The values in the row
(mPDR) correspond to the average fit parameters and the dispersion for pixels at
a distance $R$ from the center.  The actual mean values of the cloud parameters
are listed in the following row labeled ``Actual''.
We use 51 transitions in these fits and minimize the $\chi^2_{\rm red}$ by
varying four parameters, thus the DOF of the fit are 51-4 = 47. In the last
column we show the mean value of $\chi^2_{\rm red}$, which is the value of the
$\chi^2$ per degree of freedom minimizing the fits.  Smaller $\chi^2_{\rm red}$
imply better fits but not necessarily good estimates of the actual values of the
cloud parameters.
\label{tbl:paper4_fit1}}
\end{table*}

The molecular gas is indirectly affected by the FUV flux via the dust.  The dust
is heated by the FUV radiation, which in turn couples to the gas and heats it
up.
 In addition to the FUV flux, this process depends also on the gas density,
 where
it becomes very efficient for $n > 10^4$~\cmt~ and $G \gtrsim 10^3$.  In our
case, the FUV flux is unconstrained mainly because the density of the gas is
significantly lower than the critical densities of the $J > 4-3$ transitions of
CO and \thco~ and the transitions of the high density tracers, where 99\% of the
gas has a density less than $10^3$~\cmt.  These transitions are subthermally
excited and their fluxes depend strongly on density.  On the other hand, the
emission grids of these transitions as a function of $n$ and $G$ show a weak
dependence on the FUV flux. For example in \cite{mvk15-a}[Figure 7] we 
show that \gm~plays a more important role in heating the gas in the molecular 
zone compared to the heating due to the coupling of the dust to the gas.

Despite the fact that $G$ is not well constrained using the molecular lines we
have considered, which was also the case in KP15b, we did succeed to fit the
line ratios from the synthetic maps using a mechanically heated PDR.
We also learn from this exercise that it is possible to constrain $n$, \gm~and
$A_V$ with high confidence within an order of magnitude using line ratios of
high density tracers as well as CO and \thco.  

\begin{figure}[!th]
\begin{minipage}[b]{1.0\linewidth} 
\centering
\includegraphics[scale=1.0]{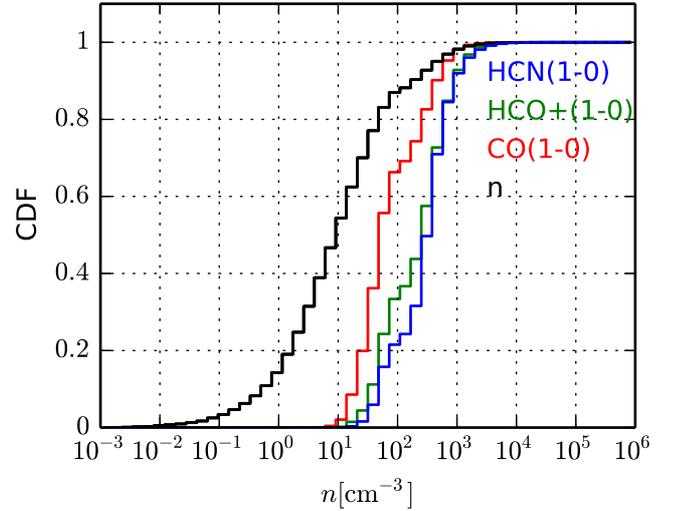}
\end{minipage}
\caption{Cumulative distribution of the densities and the luminosities of 
CO(1-0), \hcop(1-0) and HCN(1-0) for the whole galaxy.  For example, $\sim$~90\% of
 the gas particles have a density $n < 10^2$~\cmt, whereas 75\%, 40\%, 20\% of the
 emission of CO(1-0), \hcop(1-0) and HCN(1-0) emanate from gas whose density is
 lower than $10^2$~\cmt.
\label{fig:paper4_cumlum}} 
\end{figure}

It is common to have large
degeneracies when using line ratios to constrain cloud parameters.  Such
degeneracies arise mainly due to the small number of line ratios used in the
fits, while including additional line ratios reduces the degeneracy in the
parameter space.  Another reason to having degeneracies in the interpretation
of line ratios is the assumption that the molecular lines used in these 
ratios emanate on average from the same spatial region of the gas.  This is 
of-course not always true as it is evident in high resolution galactic 
studies \citep[e.g.,][]{meier12-1, meier14-1}.  The reduction in the 
degeneracies in the parameter space is illustrated in  
\cite{mvk15-a}[Figure 16], where using four independent line ratios of a sample of 
LIRGS shrinks the degeneracy to less than half a dex in the $n - G$ parameter.
Possible ways of constraining $G$ will be discussed in
Section-\ref{sec:paper4_discussion}.  The main reason why we manage to fit the
synthetic line ratios well is because the range in the densities where most of
the emission emanates is between 10 and 1000~\cmt~ (see
Figure \ref{fig:paper4_cumlum}).  This narrow range in the density is due to
the narrow density PDF of our galaxy model.  In the next section, we explore 
the effect of increasing the median and the width of the density PDF on the 
relative contribution of the whole span of densities to the flux of molecular
line emission of mechanically heated PDRs.

%-----------------------------------------------------
%-----------------------------------------------------

\section{Constraining the gas density PDF}  \label{section:constrainingpdf}

The gas density PDF is a physical property of the ISM that is of fundamental
interest because it gives information about the underlying physical processes,
such as the dynamics of the clouds and the cooling and heating mechanisms. A
log-normal distribution of the density is expected if the ISM is supersonically
turbulent \citep{Vazquez94-1, Nordlund99-1, wada01-a, wada07-1, Burkhart13-1,
hopkins13-1}, and this simple picture can explain, amongst others, the stellar
IMF, cloud mass functions, and correlation patterns in the star formation
\citep{hopkins12-1, hopkins12-2, hopkins13-1}. A log-normal distribution was
assumed in Section-\ref{sec:paper4_methods} which allows us to extrapolate the
density structure in the simulation beyond the resolution limit, in a
statistical sense.

The question remains whether it is possible to derive the properties of the
density PDF from observations. This is probably only doable for the molecular
ISM. 
There is limited scope for deriving the density PDF of the warm and cold neutral
phase of the ISM, as the HI 21 cm line is not sensitive to variations in density.
For the ionized phases, we do not expect a simple log-normal density PDF, as the
turbulence may not be supersonic and or the length scales exceed the scale
lengths in the galaxy disk.  For the colder gas, where we expect a simple
functional and relatively universal form, there is at least the prospect of
probing the density structure using the molecular line species, which come into
play at higher densities. Indeed, we may well reverse our approach taken in
Sections-\ref{subsec:paper4_constrainig}, and try to constrain the density PDF
(and the conditions in molecular clouds) using the emission of molecular
species.  Such an approach is possible because we have assembled a large
database of PDR models and resulting line emission, covering a wide range of
parameter space.

In the remainder of this section, we will explore a limited part of the
parameter space to examine the contribution from different line species for
different density PDFs under some limiting assumptions. For completeness, we
recapitulate the assumptions (some of which are implicit): 1) simple functional
forms of the density PDF, where $G$, $A_V$ and~\gm~are taken constant
\citep{Nordlund99-1}; 2) emission from each density bin in the PDF is assumed
 to come from a PDR region at that density with a fixed $A_V$, with the density dependence
coming from emission line-width and cloud size relation \citep[e.g,
][]{larson1981}.
3) chemical and thermal equilibrium is assumed for the PDR models.

We have seen in Section-\ref{subsec:paper4_emissionmaps} that the high density
tracers, in the simulation of the quiescent disk galaxy, have a limited effect
on the emission ladders. That is because the density PDF drops off very fast in
the simulation.
This ultimately derives from the exponential decay in the assumed log-normal
distribution.  For this reason, we will consider broader dispersions in the PDF.
In this paper, we adopt log-normal PDFs for the density, although a more relaxed
power law distribution could also be considered. Such a power law decay of the
density PDF is found in some models of supersonic turbulence when the effective
equation of state has a polytropic index, $\gamma$, smaller than one, and
temperatures strongly decrease with increasing density~\citep{Nordlund99-1}.

\subsection{Parameter study}

In this section, we compute the mean flux of molecular line emission emanating
from a volume of gas whose density is log-normally distributed.  In the first
part, we explore the contribution of gas, of increasing density, to the mean
flux.
 Particularly, we look for the necessary parameters of the PDF to obtain a
significant ($> 10$\%) contribution of the high density gas to the mean flux.
In the second part, we use line ratios of HCN(1-0), HNC(1-0) and \hcop(1-0) for
star-forming galaxies to constrain the parameters of the density PDF of such
systems.

In computing the mean flux for the gas in the log-normal regime, the flux
emanating from gas within a certain density range, should be weighted by the
probability of finding it within that range.  The mean flux is computed by
summing all these fluxes for all the density intervals in the log-normal regime.
 In other words, the mean flux for a volume of gas is given by:

%----------------------------
\begin{eqnarray} \label{eq:paper4_Ln}
  \bar{F} & =& N \int_{n_1}^{n_2} {{\rm F}(n) \times \rm PDF}(n)~d \ln n \\
  M & =&N \int_{n_1}^{n_2} {n \times \rm PDF}(n)~d \ln n,
\end{eqnarray}
%----------------------------

\noindent where $N$ is a normalization factor which scales the flux ($\bar{F}$)
depending on the molecular gas mass ($M$) of the region, ${\rm PDF}(n)$ is the
gas density PDF we assume for the region considered, and ${\rm F}(n)$ is the
emission flux of a given line as a function of gas density (for a fixed $G$,
\gm~and $A_V$).
Typically the bounds of the integrals are dictated by the density of the
molecular clouds, where the gas density is in the log-normal regime. A density
of 1~\cmt~ is a good estimate for the lower bound since no molecular emission is
expected for gas with $n$ less than that.  The upper bound of the integral in
Eq-\ref{eq:paper4_Ln} can be as high as $10^6$-$10^8$~\cmt.
${\rm PDF}(n)$ is a Gaussian in log scale, which decays rapidly whenever the gas
density is 1-$\sigma$ larger than the mean.  While ${\rm PDF}(n)$ is a
decreasing function of increasing density, ${\rm F}(n)$ is an increasing
function of increasing $n$.
Generally, the molecular gas mass in the central few kpc of star-forming
galaxies is on the order of $10^9-10^{10}$~\Msun~\citep{Scoville91-1,
Bryant99-1}.  This estimate, or a better one if available, can be used to
compute $N$ in Eq-\ref{eq:paper4_Ln}.  We refer to the quantity ${\rm F}(n)
\times {\rm PDF}(n)~d \ln n$ as the weighted flux.

\subsection{Weighted fluxes}

\begin{figure}[!th]
\begin{minipage}[b]{1.0\linewidth}
\centering
\includegraphics[scale=1.0]{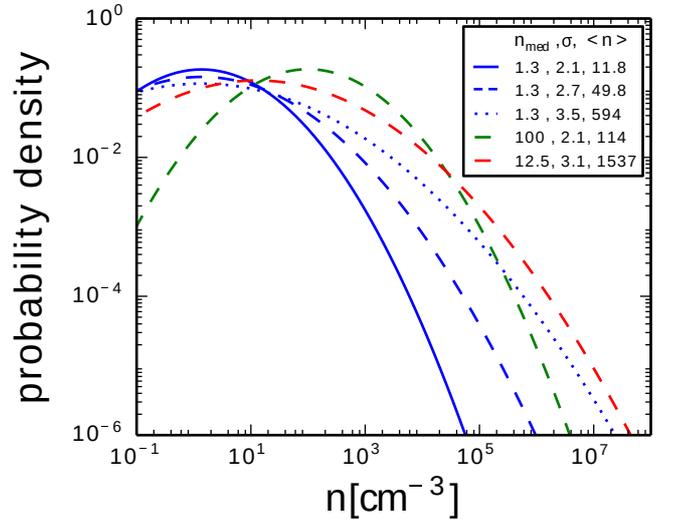}
\end{minipage} 
\caption{Effect of varying the median and the
dispersion of a density PDFs.
The blue curves represent PDFs that have the same median density as the one used
in  SPH simulation, but with increasing dispersions.  The green curve has the
same dispersion as that of the SPH simulation but with a higher median density. 
Finally, the red curve corresponds to the density PDF by Wada (2001).  The mean
densities corresponding to these PDFs are also listed in the legend.
\label{fig:paper4_syntetic_pdfs}} 
\end{figure}

\begin{figure*}[!tbhp]
\begin{minipage}[b]{1.0\linewidth}
\centering
\includegraphics[scale=0.6]{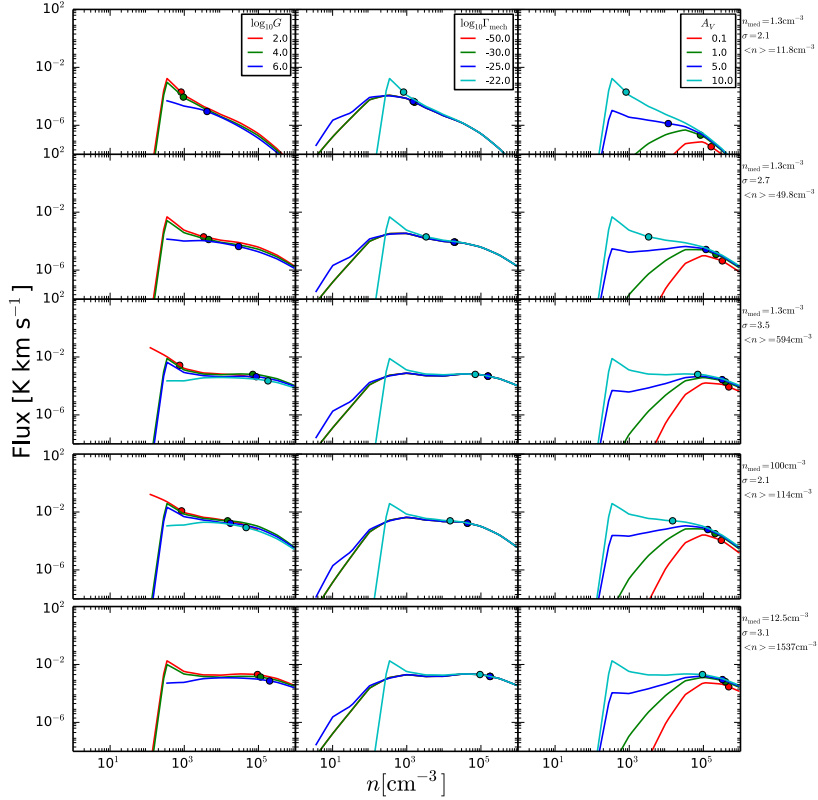}
\end{minipage} 
\caption{Weighted fluxes for HCN(1-0) for different density PDFs.  In the first
three rows the median density of the PDFs corresponds to that of the SPH
simulation, whereas the dispersion is increased from 2.1 to 2.7 to 3.5.  In the fourth
row, the median density is increased by a factor of 100 compared to that of the
SPH simulation, but the dispersion is kept fixed at 2.1.
In the last row we show the weighted fluxes for the PDF obtained by the Wada
2001 simulation.
In the first column the FUV flux is varied, from 100 times the flux at the solar
neighborhood to $10^6$\go~corresponding to the FUV flux in extreme starbursts
($A_V = 10$~mag, $\log_{10}\Gamma_{\rm mech} = -22$).  In the middle column we
vary \gm~from the heating rate corresponding to quiescent disks to rates typical
to violent starbursts with a SFR of 1000 \Msun~per year ($A_V = 10$~mag,
$\log_{10} G = 2$). In the last column the $A_V$ is varied from 0.1 to 10 mag, 
corresponding respectively to the typical values for a transition zone from
H$^+$ to H and for dark molecular clouds ($\log_{10} G = 2$,
$\log_{10}\Gamma_{\rm mech} = -22$).  The onset of emission is determined mainly
by \gm; for instance in looking at the middle column, we see that 
this onset of emission corresponds to $n \sim 10^2$~\cmt, where for lower densities
H$_2$ does not form, which is essential for other molecules such as HCN to form.
For each curve in every panel we also plot (with filled circles)
the density, $n_{90}$, where 90\% of the emission emanates from $n < n_{90}$.
For example, in the top row, even for the most intense FUV flux $n_{90} \sim
10^{3.5}$~\cmt. On the other hand, in bottom row, $n_{90} > 10^5$~\cmt.
We note that the curves corresponding to \gm~$10^{-50}$ (a pure PDR) and $10^{-30}$
\ecs~in the second column overlap.
\label{fig:paper4_syntetic_lum_PDF}} 
\end{figure*}

The median density and the dispersion of the log-normal fit in
Section-\ref{sec:paper4_methods} are $n_{\rm med} = 1.3$~\cmt~and $\sigma =
2.1$ respectively, corresponding to a mean density of $\sim 10$~\cmt.  The mean
density is much smaller than the critical densities of most of the transitions
of HCN, HNC and \hcop.  For this reason all of the emission of HCN(1-0) and
\hcop(1-0) originates from gas with $n < 10^4$~\cmt~ in
Figure \ref{fig:paper4_cumlum}.   In Figure \ref{fig:paper4_syntetic_lum_PDF},
we show the weighted fluxes of gas of increasing density for HCN(1-0).  The
fluxes are determined by computing $F$ in Eq-\ref{eq:paper4_Ln} for intervals in
$\log n$.
Similar distributions can be computed for other emission lines of high density
tracers, which are expected to be qualitatively similar.  The rows in
Figure \ref{fig:paper4_syntetic_lum_PDF} correspond to the PDFs in
Figure \ref{fig:paper4_syntetic_pdfs}.  Along the columns we vary the physical
cloud parameters over most of the expected physical conditions. $G$, \gm~and
$A_V$ are varied in the first, second and last column, respectively.
In exploring the possible ranges in $G$ and \gm~for the PDF of the SPH
simulation, we see that the peak of the emission is restricted to $n <
10^3$\cmt.  The only situation where the peak is shifted towards densities
higher than $10^3$~\cmt~ occurs when the mean $A_V$ of the clouds in the galaxy
is $\sim 1$~mag.  However, in this case, the flux would be too weak to be
observed.  Hence with the combinations of these parameters, and with such a
log-normal density PDF, it is not possible to obtain a double peaked PDF,
or a gas density distribution with significant contribution from $n >
10^4$~\cmt~ gas.  Thus, most of the emission of the high density tracers is from
gas with $n < 10^4$~\cmt.  The analogous plots of
Figures-\ref{fig:paper4_syntetic_pdfs} and \ref{fig:paper4_syntetic_lum_PDF}
where the dispersion is varied for median densities of 12.5~\cmt~ and 100~\cmt~
are shown in the appendix.

%-----------------------------------------------------
%-----------------------------------------------------

%-----------------------------------------------------
%-----------------------------------------------------

For density PDFs with broader dispersions, $\sigma =$~2.7 and 3.5, respectively,
it is possible to obtain luminosity distributions where at least 10\% of the
emission is from gas with densities $> 10^5$\cmt~ (see second, third and last
rows in Figure \ref{fig:paper4_syntetic_lum_PDF}).  In multi-component PDF fits,
the filling factor of the ``denser'' component is on the order of a few percent
(by mass or by area).  In the Figure \ref{fig:paper4_syntetic_lum_PDF}, we see
that it is necessary to have a broad dispersion in the PDF in order to have a
distribution of the luminosity where at least 10\% of it emanates from gas with
$n > 10^{5}$~\cmt.

When comparing the second and fourth\footnote{The PDF of the fourth row
corresponds to that of the simulation by Wada (2001)} rows  of the weighted
fluxes in Figure \ref{fig:paper4_syntetic_lum_PDF}, we see that they are quite
similar.
 Thus, it seems that a low median density and a broad dispersion (second row)
results in the same weighted fluxes as a PDF with a 100 times higher median
density and a narrower dispersion (fourth row).  To check for such degeneracies
and constrain the PDF parameters using molecular emission of high density
tracers, we construct line ratio grids as a function of $n_{\rm med}$ and
$\sigma$.

\begin{figure}[!tbhp]
\begin{minipage}[b]{1.0\linewidth}
\centering
\includegraphics{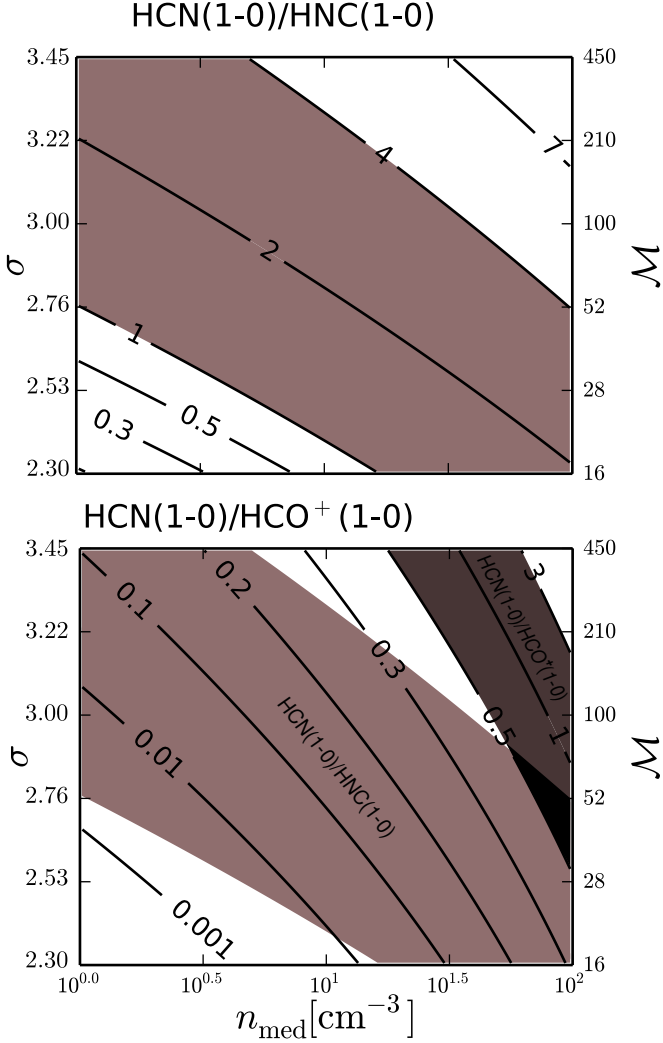}
\end{minipage} 
\caption{{\bf Top} Line ratios of HNC(1-0)/HCN(1-0) as a function of the mean
and the dispersion of the PDF. {\bf Bottom} Line ratios of \hcop(1-0)/HCN(1-0)
as a function of the mean and the dispersion of the PDF. The dark brown stripe represents
observational data. The overlapping
region of these two line ratios is colored in black in the bottom panel.  For
all the grid points a typical $A_V = 10$~mag and \gm~$=10^{-22}$~\cmt~ is used.
In both panels, Mach numbers are reported on the right axes and G is 100.
\label{fig:paper4_line_ratio_vs_density_PDF}} 
\end{figure}

It has been suggested that HCN(1-0) is a better tracer of star formation than
CO(1-0) because of its excitation properties
\citep{Gao04-1}.
As a proof of concept, in Figure \ref{fig:paper4_line_ratio_vs_density_PDF}, we
show a grid of line ratios of HCN(1-0)/HNC(1-0) and HCN(1-0)/\hcop(1-0) as a
function of the mean and the dispersion of the density PDF.
The line ratios of a sample of 117 LIRGS \cite{loenen2008}[Figures-1 and 2]
show that in most of these galaxies $1 \lesssim $~HCN(1-0)/HNC(1-0)$ \lesssim 4
$ and $0.5 \lesssim$~HCN(1-0)/\hcop(1-0)$ \lesssim 3$.  The overlapping regions
in the line ratio grids in Figure \ref{fig:paper4_line_ratio_vs_density_PDF}
correspond to  $2.54 < \sigma < 2.9$.
The width of a log-normal density PDF is related to the Mach number,
$\mathcal{M}$, in that medium via $\sigma^2 \approx \ln (1 + 4 \mathcal{M}^2 /
3)$ \citep{Krumholz07-1, hopkins12-2}.  By applying this relationship to the
range in $\sigma$ constrained by the observations we find that $ 29 <
\mathcal{M} < 77$.  This range in $\mathcal{M}$ is consistent with that of
violent starbursts that take place in extreme star-forming regions and galaxy
centers \citep{Downes98-1}.

%-----------------------------------------------------
%-----------------------------------------------------

\section{Discussion} \label{sec:paper4_discussion}

The molecular emission of star-forming galaxies usually require more than one
PDR component to fit all the transitions.  Typically, a low density PDR
components is needed to fit low-$J$ transitions, e.g., for CO and \thco, whereas
a high density $n > 10^4$~\cmt~ component is needed to fit the $J > 6-5$
transitions of these two species and the high density tracers.  In the inner $<
0.1 - 1$~kpc a mechanically heated PDR and/or an XDR might be necessary in the
presence of an AGN or extreme starbursts.
In the first part of the paper, we sampled high density gas in our model
star-forming galaxy simulation by assuming the gas density is a log-normal
function, in order to account for the emission of the high density gas that was
missing in the model galaxy due to resolution constraints.  Since the dispersion
of the density PDF of the model galaxy is narrow with $\sigma = 2.1$,
corresponding to a Mach number $\mathcal{M} \sim 10$, the gas with density $n >
10^4$~\cmt~ contributes $< 1\%$ of the total luminosity of each line.
Consequently we were able to recover $n$, \gm~and $A_V$ of the gas parameters
within 2 kpc from line ratios of CO, \thco, HCN, HNC and \hcop, reasonably well
using a one component mechanically heated PDR model.
The FUV flux was constrained less accurately, since \gm~is a dominant heating
term at $A_V \gtrsim 1$~mag, where most of the molecular emission emanates. We
have seen in \cite{mvk15-a} that in the non-LTE regime, the molecular emission
is almost independent of the FUV flux.  This is not the case for $n \gtrsim
10^4$~\cmt, however for such high densities the line ratio grids depend strongly
on $G$, thus whenever the mean gas density is $> 10^4$~\cmt~and $G > 10^4$, e.g.
in ULIRGS, we expect to constrain the FUV flux with high certainty.  In this
case it might be necessary to model the emission with more than one PDR
component as was done by \cite{2014A&A...568A..90R}. In this paper, the CO
ladder of the ULIRG Arp 299 was fit using three PDR models and the FUV flux was
well constrained only for the densest component with $n = 10^{5.5}$\cmt~(See
\cite{2014A&A...568A..90R}[Figure 5]).
Another possible way of constraining $G$ is using diagnostic line ratios of
atomic fine-structure lines.  Since atomic fine-structure emission originates
from $A_V< 1$~mag, it depends strongly on $G$.  Consequently, these line ratios
show a strong dependence on $G$ as well \citep[see review by][and references
therein]{2013RvMP...85.1021T}.
This is valid even for the most extreme \gm~rates (see Fig-A3 by
\cite{mvk15-a}).

Modeling the ISM as discrete components is not a realistic
representation, especially in ULIRGS, since the gas in such environments is
expected to be distributed log-normally and in some cases the distribution could
be a power-law (depending on the adiabatic index of the equation of state).

We have studied the contribution of the density function to the mean flux for
different parameters defining a log-normal probability functions and showed that
a broad dispersion is required to obtain significant emission for the $n >
10^4$~\cmt~ clouds.  The main advantage of such modeling is in interpreting
un-resolved observations of star-forming galaxies where FUV heating and
mechanical heating may play an important role. This approach is more appropriate
than fitting the observations with one or more PDR components, since in such
modeling the gas is assumed to be uniformly distributed with discrete densities.
 The number of free parameters arising from multi-component fits would be much
higher than fitting the parameters of the gas density PDF, which has just two. 
This reduces the degeneracies and gives us information on the gas density PDF
and about the turbulent structure and the Mach number, which is directly related
to the density PDF and the mechanical heating rate.
The main caveat in our approach is the assumption that the density PDF for $n >
10^{-2}$~\cmt~is a log-normal function and that the mechanical heating rate used
in the PDR models is independent of the width of density PDF, A$_V$, $G$ and the
line-width.  The reason for adopting this assumption is based on the fact that a
relationship between gas density, and consequently the PDF, and the mechanical
heating rate is lacking \citep{wheeler80-1,scilich96-1,freyer03-1,freyer06-1}.
Ultimately the mechanical heating rate derives from the cascade of turbulence to
the smallest scales due to a supernova event, where typically an energy transfer
efficiency of 10\% is assumed \citep{loenen2008}.
The outcome of recovering cloud parameters by independently varying them in the
fitting exercise is a good step towards probing possible relationships between
these parameters and mechanical heating.
We have demonstrated the possibility of constraining the density PDF using line
ratios of HCN(1-0), HNC(1-0) and \hcop(1-0), where the derived Mach number is
consistent with previous predictions.  Grids of diagnostics involving other
molecular species can also be computed \cite[see reviews by][and references
therein]{wolfire2011-1, bergin2011-1,aalto2014-1} in order to constrain the
properties of the PDF, but that is beyond the scope of this paper.
Moreover, we have used the flux ratios of molecular lines as diagnostics, but it
is also possible to use the ratio of the star formation rate (SFR) to the line
luminosity to constrain the PDF as is done by \cite{Krumholz07-1}.
This is in fact quite interesting, since the SFR can be related to the
mechanical heating rate as was done by \cite{loenen2008}.  By doing so, a
tighter constraint on the mechanical heating rate can be imposed, instead of
considering it as a free parameter as we have done in our fitting procedure.

\section{Summary and Conclusion}

We have constructed luminosity maps of some molecular emission lines of a
disk-like galaxy model.  These emission maps of CO, \thco, HCN, HNC, and
\hcop~have been computed using subgrid PDR modeling in post-processing mode.
Because of resolution limitations, the density of the simulation was restricted
to $n < 10^4$~\cmt.  We demonstrated that the density PDF is log-normal for $n >
10^{-2}$~\cmt. Most of the emission of the high density tracers emanates from
the gas with densities $\sim 10^2$~\cmt~ for quiescent galaxies, which is at
least 1000 times lower than the critical density of a typical high
density tracer.  We attribute this to the fact that the dispersion of the PDF is
narrow, and thus the probability of finding dense gas is low.

The main findings of this paper are:

\begin{itemize}
\item It is necessary to have a large dispersion in the density PDF ($\sigma >
2.7$) in order to have significant emission of high density tracers from $n >
10^4$~\cmt~ gas.
\item It is possible to constrain the shape of the PDF using line ratios of high
density tracers.
\item Line ratios of HCN(1-0), HNC(1-0), and \hcop(1-0) for star-forming galaxies
and starbursts support the theory of supersonic turbulence.
\end{itemize}

A major caveat for this approach is the assumption concerning the thermal and
the chemical equilibrium.  Care must be taken in interpreting and applying such
equilibrium models to violently turbulent environments such as starbursting
galaxies and galaxy nuclei.
Despite the appealing fact that the line ratios obtained from the example we
have shown in Figure \ref{fig:paper4_line_ratio_vs_density_PDF} favor high Mach
numbers ($ 29 < \mathcal{M} < 77$) consistent with previous prediction of
supersonic turbulence in starbursts, a time-dependent treatment might be
essential.

%-----------------------------------------------------
\begin{acknowledgements}
M.V.K is grateful to Alexander Tielens for useful comments and suggestions. 
M.V.K also thanks the anonymous referee and Volker Ossenkopf for their
critical reviews on the manuscript that helped improve it significantly.
\end{acknowledgements}

\bibliographystyle{aa}
\bibliography{main}

%-----------------------------------------------------
%-----------------------------------------------------
\clearpage
\begin{appendix} 
%%% this is necessary to get the 'A' before the number of the figure
\end{appendix}

\begin{appendix}{Appendix}
\label{app:a}

%-----------------------------------------------------
\begin{figure}[!tbhp]
\begin{minipage}[b]{1.0\linewidth}
\centering
\includegraphics[scale=1.0]{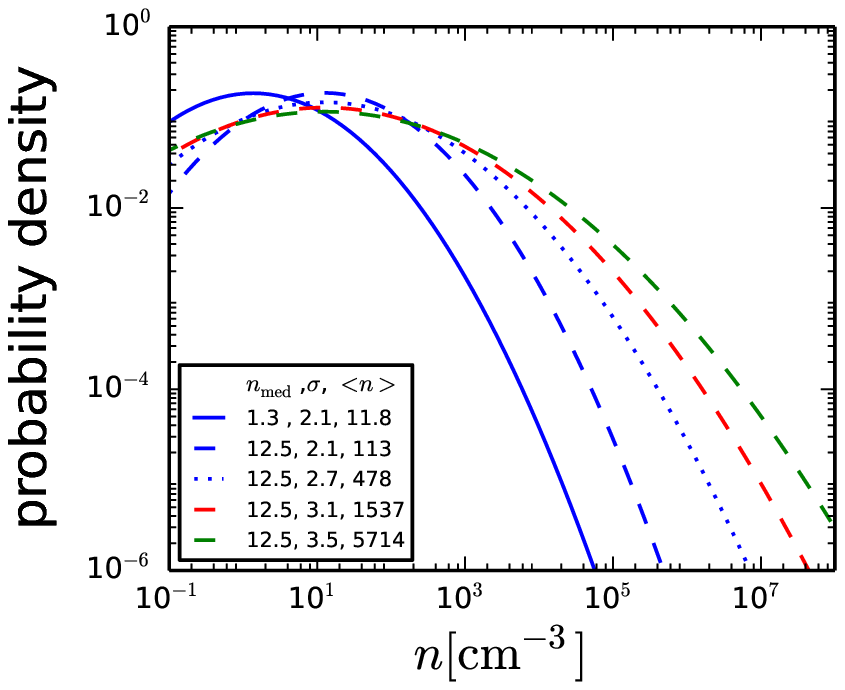}
\includegraphics[scale=1.0]{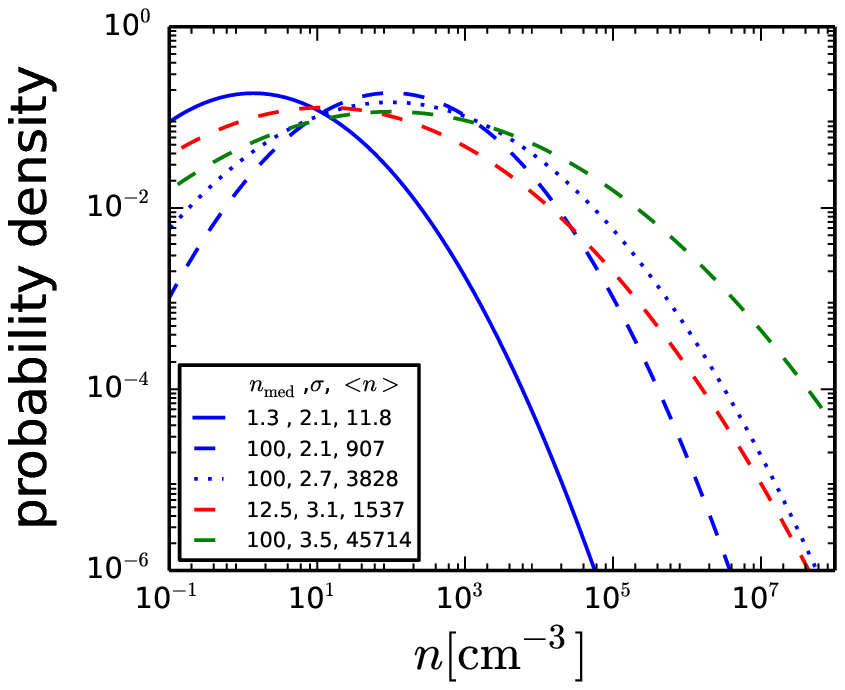}
\end{minipage} 
\caption{Analogous to Figure \ref{fig:paper4_syntetic_pdfs} where the
dispersion is varied for a fixed median density of 12.5~\cmt~ in the left panel
and 100~\cmt~ in the right panel.  In both cases the density PDF of the SPH
simulation and the Wada 2001 simulation are also shown for reference and
comparison.\label{fig:paper4_syntetic_pdfs_app}}
\end{figure}
%-----------------------------------------------------

%-----------------------------------------------------
\begin{figure*}[!tbhp]
\begin{minipage}[b]{1.0\linewidth}
\includegraphics[scale=0.7]{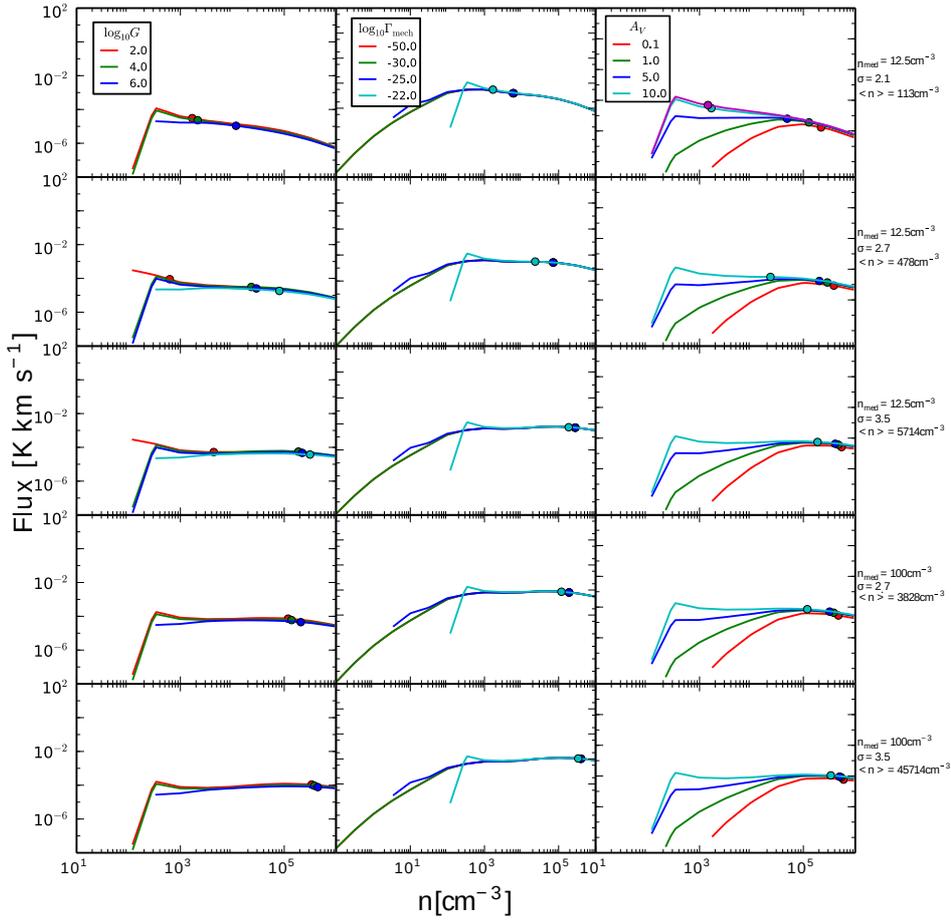}
\end{minipage} 
\caption{Weighted fluxes of the PDFs of
Figure \ref{fig:paper4_syntetic_pdfs_app} which are not shown in
Figure \ref{fig:paper4_syntetic_lum_PDF}
\label{fig:paper4_syntetic_lum_PDF_app}}
\end{figure*}
%-----------------------------------------------------

\end{appendix}

\end{document}